\documentclass[prb,nopacs,twocolumn,preprintnumbers,amsmath,amssymb]{revtex4}

\usepackage{epsfig}
\usepackage{graphicx}
\usepackage{dcolumn}
\usepackage{color}
\usepackage{enumerate}
\usepackage{bm}
\newcommand{\be}{\begin{equation}}
\newcommand{\ee}{\end{equation}}
\newcommand{\red}{\color {red}}

\begin{document}

\title{Chiral charge pumping in graphene deposited on a magnetic insulator}

\author{Michael Evelt,$^1$ Hector Ochoa,$^2$ Oleksandr Dzyapko,$^1$  Vladislav E. Demidov,$^1$  Avgust Yurgens,$^3$  Jie Sun,$^3$ Yaroslav Tserkovnyak,$^2$ Vladimir Bessonov,$^4$ Anatoliy B. Rinkevich,$^4$ and Sergej O. Demokritov$^{1,4}$}

\affiliation{$^1$Institute for Applied Physics and Center for nanotechnology, University of Muenster, 48149 Muenster, Germany\\
$^2$Department of Physics and Astronomy, University of California, Los Angeles, California 90095, USA\\
$^3$Department of Microtechnology and Nanoscience-MC2, Chalmers University of Technology, SE-41296, Gothenburg, Sweden\\
$^4$Institute of Metal Physics, Ural Division of RAS, Yekaterinburg 620041, Russia}

\begin{abstract}
We demonstrate that a sizable chiral charge pumping can be achieved at room temperature in graphene/Yttrium Iron Garnet (YIG) bilayer systems. The effect, which cannot be attributed to the ordinary spin pumping, reveals itself in the creation of a dc electric field/voltage in graphene as a response to the dynamic magnetic excitations (spin waves) in an adjacent out-of-plane magnetized YIG film.  We show that the induced voltage changes its sign when the orientation of the static magnetization is reversed, clearly indicating the broken spatial inversion symmetry in the studied system. The strength of effect shows a non-monotonous dependence on the spin-wave frequency, in agreement with the proposed theoretical model.
\end{abstract}

\maketitle

\section{Introduction}


The generation of spin-polarized currents is of technological interest in order to transmit information encoded in the spin degrees of freedom. A spin current can be injected into a nonmagnetic conductor by spin pumping.\cite{pumping_theo1,pumping_theo2} In this process, the reservoir of the angular momentum is a ferromagnet where a ferromagnetic resonance (FMR) is excited by a microwave field. The spin pumping without net charge transport stems from the exchange of angular momentum between the itinerant spins in the metal and the collective magnetization dynamics in the adjacent ferromagnet. The dynamical generation of spin currents is of special relevance since alternative methods based on driving an electrical current through the interface are limited by the conductance mismatch.\cite{mismatch} The spin pumping-induced currents have been detected as the long-ranged dynamic interaction between two ferromagnets separated by a normal metal\cite{pumping_exp1,pumping_exp2} or as a dc-voltage signal in bilayer systems.\cite{pumping_exp3} In the case of nonmagnetic metals with a strong spin-orbit coupling, the spin current through the interface may engender a charge current along the metallic layer as result of the inverse spin-Hall effect.\cite{spin-charge1,spin-charge2}  As the thickness of the metallic layer is reduced, additional interfacial effects must be considered. The different contributions may be identified by analyzing the symmetries of the voltage signal with respect to different adjustable parameters, such as the orientation of the saturated magnetization controlled by the applied static field.\cite{bilayer} Additionally, we can use propagating spin waves instead of FMR and vary the direction of propagation.

Among other materials, the unique candidate for investigations of interface-induced phenomena is single layer graphene (SLG). Due to its unique mechanical, optical, and electronic properties, graphene has attracted enormous attention since its discovery in 2004.\cite{10,12} Nowadays, one can produce large-area high-quality SLG by using, e.g., chemical vapor deposition on metal catalysts.\cite{6,7,8} For the observation of spin pumping effects, SLG should be brought in contact with a magnetic material. 
Yttrium Iron Garnet (YIG) holds a special place among all magnetic materials. Specifically, it shows an unprecedentedly small magnetic damping resulting in the narrowest known line of the FMR and enabling propagation of spin waves over long distances.\cite{13,14} Due to these unique characteristics, YIG films have been recently considered as a promising material for spintronic and magnonic applications.\cite{15,16} By combining YIG and SLG in a bilayer, one obtains a unique model system for investigations of the spin-wave induced interfacial effects.

Here, we report the experimental observation of a chiral pumping effect in out-of-plane magnetized YIG/SLG bilayer and show that spin waves propagating in the YIG film induce a sizable electric voltage in the adjacent SLG. We study the symmetry of the observed phenomenon by reversing the static magnetic field and the direction of spin-wave propagation and find that the corresponding changes in the sign of the electric voltage clearly indicate a broken reflection symmetry about the plane normal to the interface.
We associate this symmetry breaking with the presence of screw dislocations in the crystalline lattice of YIG, which are expected to have a strong effect on the two-dimensional electron gas in SLG. We present a theoretical model, which shows that such crystalline defects can result in a chiral pumping effect with the observed symmetry. It is worth mentioning that the conventional spin pumping-induced voltages obtained for in-plane magnetized films\cite{17,9,18} are rooted in the reflection symmetry breaking about the plane of the interface. They present different symmetries and are therefore unrelated to the signal reported here. Moreover, in agreement with the experiment, our model predicts a non-monotonic dependence of the induced voltage on the spin-wave frequency.

The structure of the manuscript is the following: We present the microwave and voltage measurements in Sec.~\ref{sec:experiments}. Special attention is paid to the inversion symmetry breaking, confirmed by further FMR experiments in YIG, revealing the chiral nature of the effect. The theoretical model is discussed in Sec.~\ref{sec:theory}. Technical details of the model are set-aside in the Appendix. We summarize our findings in Sec.~\ref{sec:conclusions}.

\section{Ferromagnetic resonance experiments}

\label{sec:experiments}

\subsection{Sample preparation and characterization}

The experimental layout is shown in Fig. 1(a). Graphene was placed on top of a monocrystalline YIG film of 5.1 $\mu$m thickness grown by means of liquid phase epitaxy on 0.5 mm thick gallium gadolinium garnet (GGG) substrate. The saturation magnetization of the YIG film defining the saturation field in the out-of-plane geometry was $4\pi M_S=1.75$ kG. The SLG sample was grown on a 50 $\mu$m thick 99.99\% pure copper foils in a cold-wall low-pressure CVD reactor (Black Magic, AIXTRON). After the Cu foil had been cleaned in acetone and then shortly etched in acetic acid to remove surface oxide, it was placed on a graphitic heater inside the reactor and annealed at 1000 $^{\circ}$C during 5 min in a flow of 20 sccm H$_2$ and 1000 sccm Ar. Then, a flow of pre-diluted CH$_4$ (5\% in Ar, 30 sccm) was introduced during 5 min to initiate the growth of graphene. After the growth, the gas flow was shut down, the system evacuated to $<0.1$ mbar and cooled down to room temperature.\cite{28} SLG was transferred to YIG/GGG by using the bubbling delamination technique\cite{29,30} and lithographically patterned into a multi-terminal Hall-bar structure with Au(100 nm)/Cr(5 nm) pads at the edges. The lateral dimensions of the sample were 1.5 by 25 mm. The Hall-effect mobility of the resulting devices was found to be in the range from 600 to 1300 cm$^2$ (V$\cdot$s)$^{-1}$ at T=10 K.  The local resistance maximum at zero magnetic field $H_0\approx0$, usually ascribed to the weak localization effect, allowed for an estimation of the phase coherence length $L_{\phi}\sim 100$ nm at the same temperature. {\red 
} 

\subsection{Microwave and voltage measurements}

\begin{figure}
\includegraphics[width=0.9\columnwidth]{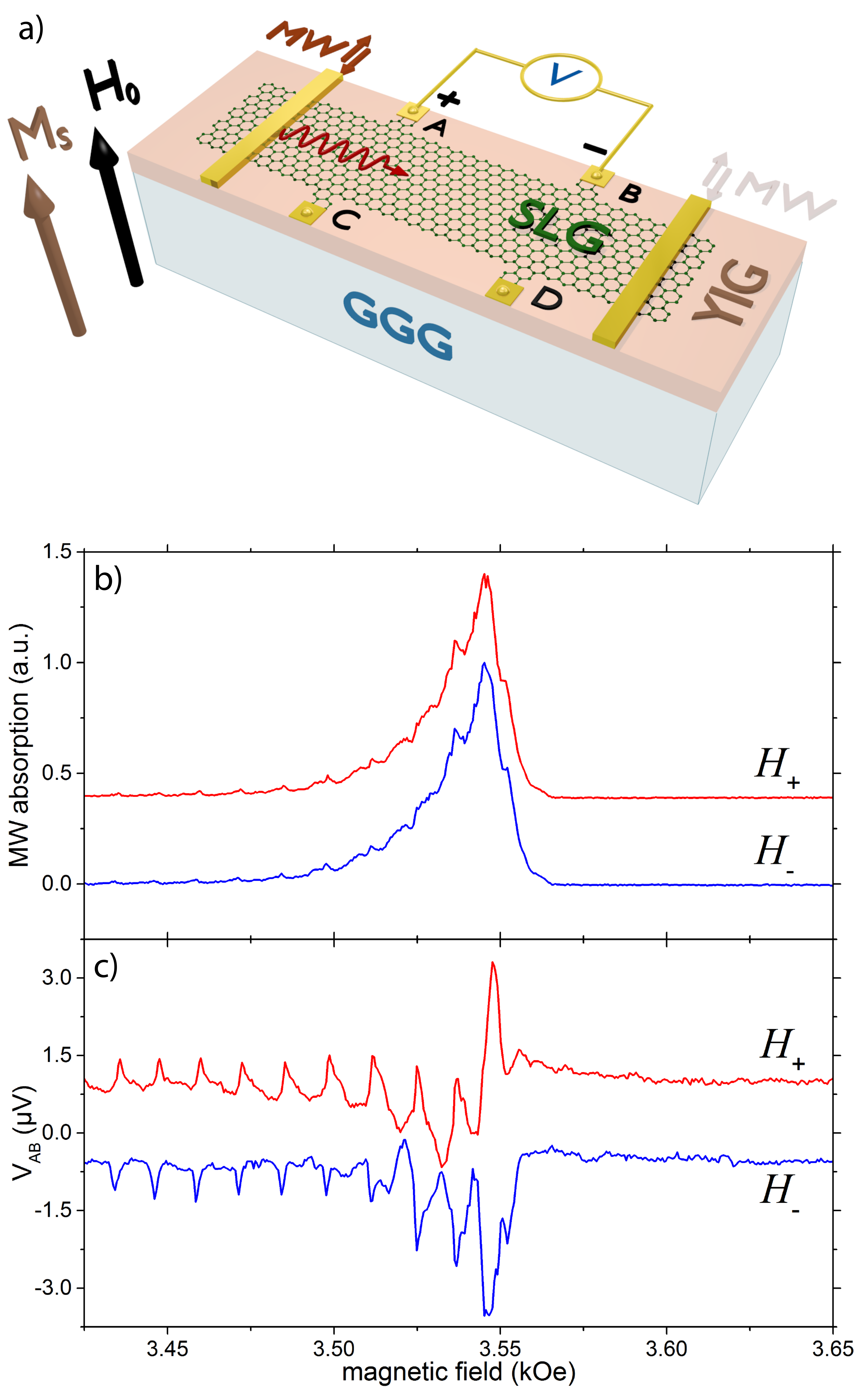}
\caption{Chiral charge pumping in YIG/SLG. a) Schematic of the experiment. Single layer graphene (SLG) is transferred to yttrium-iron-garnet (YIG) film grown on gallium gadolinium garnet (GGG) substrate. A, B, C, D are gold contact patches for electric measurements. The system is magnetically saturated normally to the sample plane by means of an applied magnetic field $H_0$. Spin waves are excited by microwave (MW) current flowing in one of the two narrow antennae, producing a local MW field. b) Field dependence of the MW absorption for the \textit{up} ($H_+$) and \textit{down} ($H_-$) orientation of the static field. Different peaks on the dependences indicate excitation of different non-uniform spin-wave modes. Note the similarity of the two curves. c) Field dependence of the dc-voltage detected between the pads A and B for the \textit{up} ($H_+$) and \textit{down} ($H_-$) orientation of the applied field. Note different signs of the voltages for the two orientation of the field. The red curves in (b) and (c) are shifted vertically for the sake of clarity.}
\label{Fig1}
\end{figure}

\begin{figure*}
\includegraphics[width=0.9\textwidth]{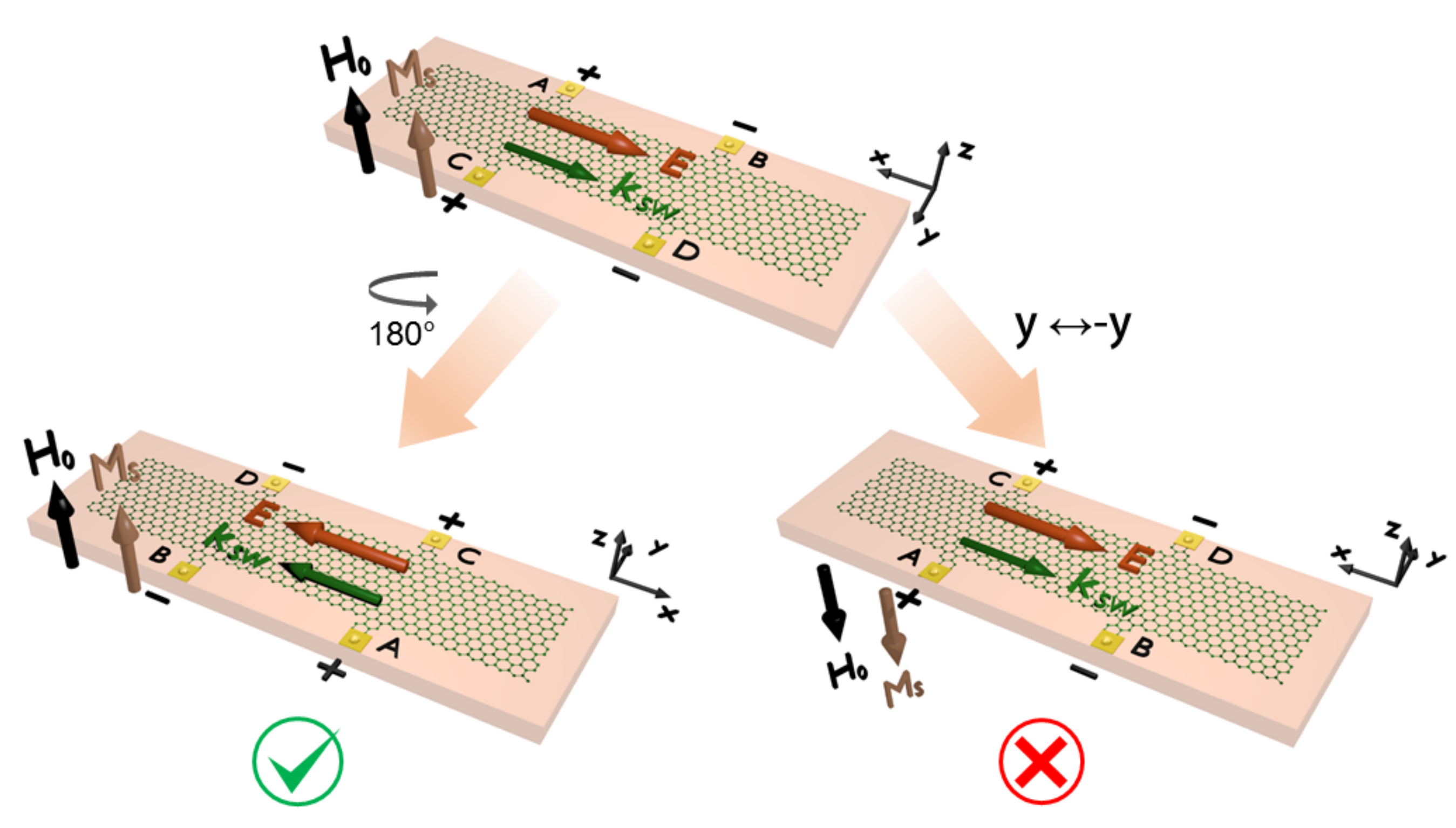}
\caption{Symmetry properties of the observed effect. Both the induced electric field $E$ and the wavevector of the spin wave $\bm{k}_{\textrm{sw}}$ are in the plane of the sample, while the applied field $H_0$ and the YIG-magnetization $M_S$ are oriented normally to the sample plane. The $+$ and $-$ signs indicate the polarity of the measured dc-voltage. Rotation of the system by 180$^{\circ}$ around the $z$-axis reverses the directions of both $E$ and $\bm{k}_{\textrm{sw}}$ in agreement with the experiment. Mirroring with respect to the vertical plane ($y\rightarrow-y$) does not change $E$ and $\bm{k}_{\textrm{sw}}$, while $H_0$ and $M_S$ are reversed, in contrast to the experimental findings.}
\label{Fig2}
\end{figure*}

The YIG/SLG sample was built onto a massive metal holder whose temperature was controlled and kept constant at $295 \pm 0.02$ K. The holder with the sample was placed between the poles of an electromagnet, which creates a static magnetic field $H_0$ with the inhomogeneity below $2\times10^{-5}$ over 1 mm$^{3}$. In all described experiments we kept $H_0> 2$ kOe ensuring the collinear orientation of $H_0$ and $M_S$. To excite spin waves in the YIG layer, broadband microstrip lines ending with 50 $\mu$m wide antennas were used, which were attached to the sample. Microwave (MW) field with a fixed frequency in the range $f=3-8$ GHz was applied to the antennas. A directional coupler was used to monitor the reflected MW power, allowing for the determination of the resonance absorption in the sample due to the excitation of the spin waves. By keeping $f$ constant and varying $H_0$, we recorded $P_{abs}\left(H_0\right)$ -- curves characterizing the field-dependent excitation of spin waves in YIG, see Fig.~\ref{Fig1}~b). The curves exhibit several maxima corresponding to the FMR, higher-order standing, as well as propagating spin waves in YIG.

Simultaneously with the microwave measurements, we recorded the dc-voltage between the electrodes. A lock-in technique was applied in order to detect the dc-voltage caused by the spin waves propagating in the YIG film. Microwaves were modulated by a square wave with the repetition frequency of 10 kHz and the voltage difference between different leads was measured by the lock-in amplifier. During the measurements, the static magnetic field was swept and the dc-voltage along with the reflected MW power signal as a function of $H_0$ were recorded. Figure 1(c) shows a typical dependence of $V_{\textrm{AB}}\left(H_0\right)$ -- the voltage between the electrodes A and B. Comparing Figs. 1(b) and 1(c), one sees that the dc-voltage in SLG is induced at the same values of $H_0$, where the efficient excitation of spin waves takes place. This clearly indicates its direct connection to the high-frequency magnetization precession in YIG.

\subsection{Symmetry breaking}

The most striking feature of the observed phenomenon is its unusual symmetry with respect to the inversion of $H_0$. As seen from Fig. 1(c), the detected voltage changes its sign if the orientation of $H_0$ (and $M_S$) is reversed. We emphasize that this inversion is not accompanied by visible changes in the microwave absorption curve (Fig. 1(b)). Further measurements show that the induced voltage is the same for both sides of the sample, $V_{\textrm{AB}} = V_{\textrm{CD}}$. It is proportional to the intensity of the spin wave (the absorbed microwave power) and changes sign if the direction of the spin-wave propagation is reversed.

To obtain better insight into the underlying physics, we consider the symmetry of the phenomenon in more detail. As shown in Fig. 2, the induced electric field $E$ is oriented in the sample plane parallel to the direction of the spin-wave propagation indicated in Fig. 2 by a vector $\bm{k}_{\textrm{sw}}$, whereas the magnetization/magnetic field is perpendicular to the sample plane. The $+$ and $-$ signs indicate the polarity of the measured dc-voltage at a given orientation of the magnetization. As seen in Fig. 2, rotation by 180$^{\circ}$ around the normal to the film plane reverses both the direction of the spin-wave propagation and that of the electric field, whereas the direction of the magnetization stays unchanged, all in agreement with the experimental findings. However, mirroring of the system with respect to the vertical plane parallel to the direction of the spin-wave propagation ($y\rightarrow-y$) does not change neither the direction of the spin-wave propagation nor that of the electric field, since they are real vectors, whereas the magnetization, which is an axial vector, is reversed. We emphasize that this contradicts to the experiment (see Fig. 1 c)). In other words, the phenomenon responsible for the appearance of the voltage in the YIG/SLG bilayer requires that the inversion symmetry is broken. 
More specifically, the observed effect would be forbidden if the clockwise and counter-clockwise directions in the sample plane were equivalent, which reveals its chiral nature.

\begin{figure}
\vspace{-0.8cm}
\includegraphics[width=1\columnwidth]{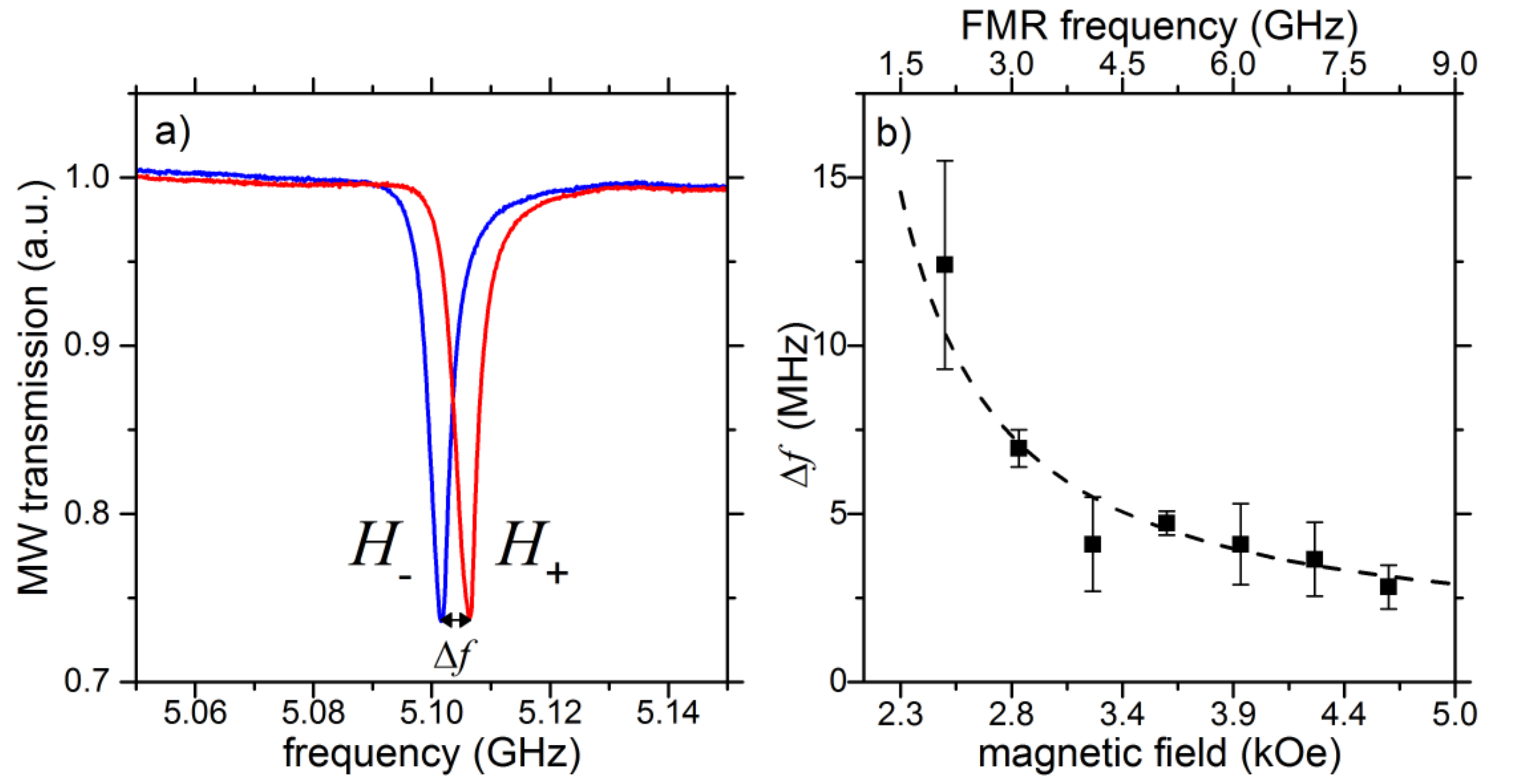}
\caption{Ferromagnetic resonance (FMR) in a YIG film. a) FMR curves measured at a constant $H_0$ for the \textit{up} ($H_+$) and \textit{down} ($H_-$) orientation of the field. Note a difference in the FMR frequencies $\Delta f$. b) Field dependence of $\Delta f$. The dash line is a guide for the eye.}
\label{Fig3}
\end{figure}

To check whether the observed non-equivalence is a characteristic feature of the YIG film itself, we performed high-precision FMR measurements on YIG films of small lateral dimensions (5.1 $\mu$m thick circle with the 0.5 mm radius and 6 $\mu$m thick $0.5\times0.5$ mm$^2$ square) without SLG. In these experiments, the FMR in the YIG sample was excited by a quasi-uniform dynamic magnetic field. The entire test device was mounted on a non-magnetic rotatable sample holder, which was placed in a static magnetic field with a homogeneity better than 0.1 Oe, with the $x$-axis being the rotation axis of the holder. Note here that the actual rotation axis (the $x$-axis) in the described experiments differs from the rotation axis of the thought experiment (the $z$-axis) shown in Fig. 2. The sample holder was designed to keep the position of the sample constant with an accuracy of 0.2 mm over the entire rotation. By sweeping the microwave frequency at a fixed magnetic field, the transmission coefficient $T=P_{out}/P_{in}$ was measured as a function of the frequency. Then, the holder with the sample was rotated by 180$^{\circ}$, and the measurements were repeated.  The results of the measurements for the two orientations of the sample were averaged over several measurement cycles. The obtained FMR curves for the directions of the static magnetic field parallel $(H_+)$ and antiparallel $(H_-)$ to the normal to the film surface demonstrating a clear frequency difference of $\Delta f=5$ MHz are shown in Fig.~\ref{Fig3}~(a).  The value of $\Delta f$ is significantly larger compared to the estimated uncertainty of 0.5 MHz originating from the inhomogeneity of the static field and the stray fields of the sample holder. The measurements were repeated for different values of the applied magnetic field resulting in the field dependence of $\Delta f$ shown in Fig.~\ref{Fig3}~(b). One clearly sees that $\Delta f$ systematically increases as $H_0$ approaches $4\pi M_S=1.75$ kG. Qualitatively similar results were obtained for different samples except that the maximum values $\Delta f$ were found to vary from 6 to 13 MHz.

The results presented in Fig. 3 show that the frequency of the FMR in YIG films depends on whether the static magnetic field is parallel or antiparallel to the normal to the film surface $\hat{\mathbf{z}}$, i.e. it depends on the direction (clockwise or counter-clockwise with respect to $\hat{\mathbf{z}}$) of the magnetization precession. This indicates that the inversion symmetry is broken, since the clockwise and the counter-clockwise directions in the plane of the film are not equivalent. Although the microscopic origin of this breaking is not yet fully clear, it should be connected with the defects in the crystallographic structure of YIG, since the ideal high-symmetry cubic structure of YIG is definitely incompatible with this symmetry breaking. It is known that the dominating defects in high-quality epitaxial YIG films are growth dislocations.  Their typical lateral density of about 10$^8$ cm$^{-2}$ is connected with the typical misfit between YIG and GGG lattices of $10^{-3}$ (Refs.~\onlinecite{13,19}). It is also known that such dislocations strongly influence the magnetic dynamics in YIG.\cite{20} Possible mechanisms of the defect-mediated symmetry breaking can include a growth-induced misbalance between clockwise and counter-clockwise screw dislocations and effects similar to the antisymmetric surface Dzyaloshinskii-Moriya-like interactions\cite{21} in combination with a dislocation.
 
We emphasize that, although the symmetry breaking is present in YIG films without SLG on top, its influence on the magnetization dynamics is very small and can only be detected in high-precision measurements. This is not surprising, since the symmetry breaking appears to be a surface phenomenon, which has vanishing influence on the magnetization in the bulk of the film. On the contrary, this symmetry breaking is expected to have significant influence on conduction electrons in SLG placed on the surface of the YIG film.

\section{Phenomenological model}

\label{sec:theory}

Next, we provide a phenomenological model for the observed effect based on the existence of a finite density of screw dislocations in YIG. The voltage signal is related with the electromotive forces induced by the magnetization dynamics. A screw dislocation creates a distortion of the graphene lattice as shown in Fig.~\ref{Fig4}~(a), which couples to the electron spin through the spin-orbit interaction. As a result, the exchange field seen by the itinerant electrons is tilted with respect to the magnetization in YIG, Fig.~\ref{Fig4}~(b), modifying the longitudinal response in the non-adiabatic regime. We emphasize that this mechanism must be taken as a suggestive explanation for the observed phenomenon since there is no direct observation of the proposed mechanical deformations.

\subsection{Hamiltonian}

We assume that the exchange interaction couples the spins of itinerant electrons of SLG to the localized magnetic moments in YIG. The Hamiltonian reads as\begin{align}
\mathcal{H}=\hbar\, v_F\,\mathbf{\Sigma}\cdot\mathbf{p}+\Delta_{\textrm{ex}}\text{ }\bm{m}\left(t,\mathbf{r}\right)\cdot\mathbf{s}+\mathcal{H}_{\textrm{s-l}}.
\label{eq:H}
\end{align}
The first term is the Dirac Hamiltonian describing electronic states with momentum $\mathbf{p}$ around the two inequivalent valleys $\mathbf{K}_{\pm}$ in graphene, where $\mathbf{\Sigma}=\left(\pm\sigma_x,\sigma_y\right)$ is a vector of Pauli matrices associated to the sub-lattice degree of freedom of the wave function and $v_F$ is the Fermi velocity. The second term corresponds to the exchange coupling, $\Delta_{\textrm{ex}}$, where $\mathbf{s}=\left(s_x,s_y,s_z\right)$ are the itinerant spin operators and $\bm{m}\left(t,\mathbf{r}\right)$ is a unit vector along the local magnetization in YIG. The last term is a spin-lattice interaction that couples the mechanical degrees of freedom with the spins of electrons.

\begin{figure}
\vspace{-1.cm}
\includegraphics[width=1\columnwidth]{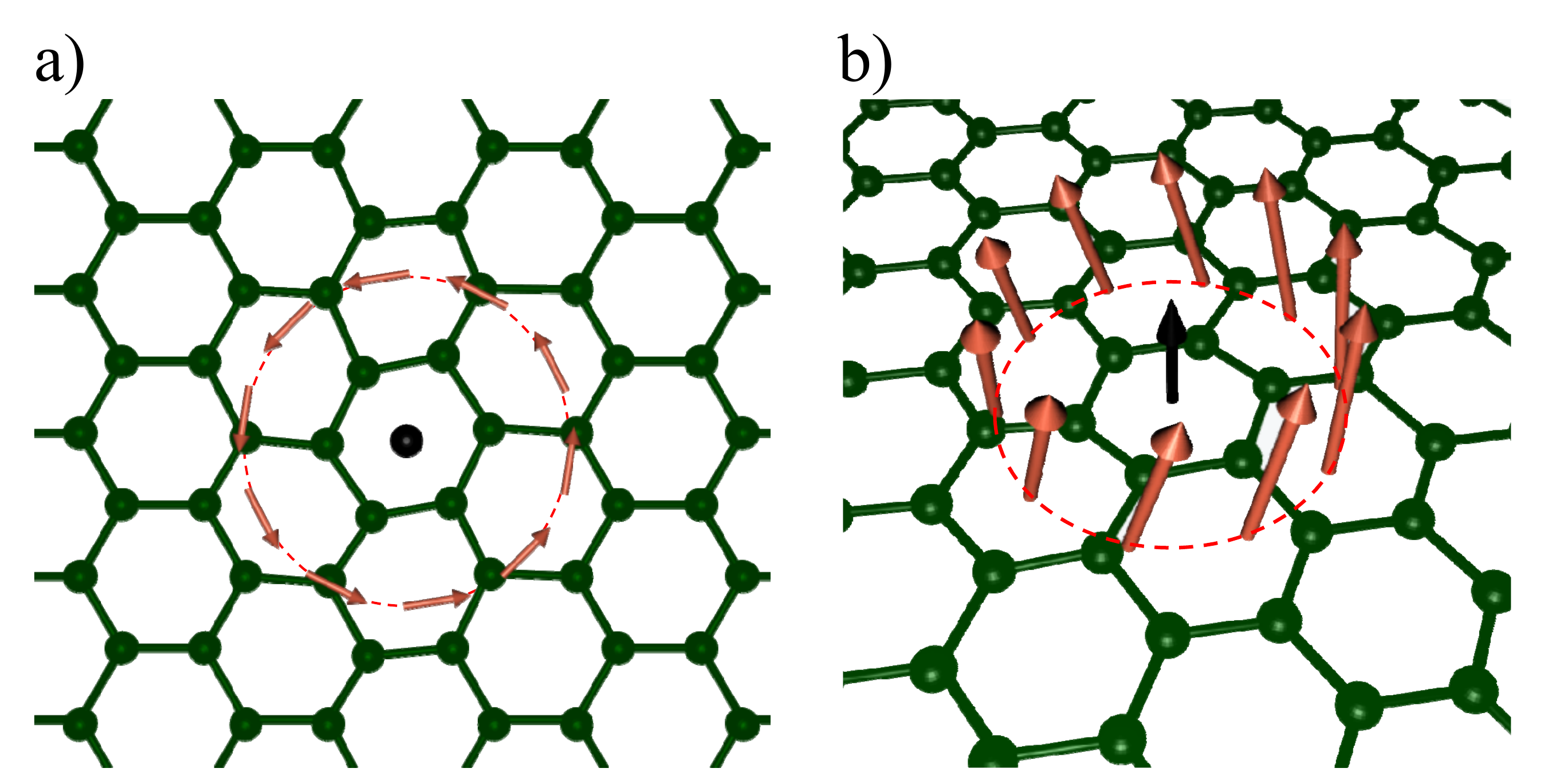}
\caption{Distorted SLG near a dislocation in a YIG-film. a) Top view of the distortion induced by a screw dislocation (the dislocation line is represented by the black dot). The red arrows illustrated the in-plane component of the effective exchange field. b) Isometric illustration of the tilt of the effective exchange field (red arrows) along a path encircling the dislocation line (black arrow). The original field points along the normal to the sample plane, and the tilting is generated by the spin-lattice coupling.}
\label{Fig4}
\end{figure}

As it is well known,\cite{12} graphene can support large mechanical distortions that change dramatically the dynamics of Dirac electrons. When an SLG is placed on top of disturbed YIG, the carbon atoms near a screw dislocation can be expected to follow its distorted profile, which introduces a torsion in the honeycomb lattice, as represented in Fig.~\ref{Fig4}~(a). This deformation couples to the electron spin as \begin{align}
\label{eq:s-l}
\mathcal{H}_{\textrm{s-l}}=\Delta_{\textrm{so}} (\partial_x u_y-\partial_y u_x )  (\sigma_x s_x+\sigma_y s_y),
\end{align}
where $\partial_x u_y-\partial_y u_x$ is the anti-symmetrized strain tensor -- $\bm{u}=\left(u_x,u_y\right)$ represents the displacements of the carbon atoms -- that parameterizes the torsion of the graphene lattice and can on average be related to the density of screw dislocations. The form of this coupling is dictated by the C$_{6v}$ symmetry of the honeycomb lattice, and its microscopic origin resides in the hybridization of $\pi$ and $\sigma$ orbitals of carbon atoms, which enhances the spin-orbit effects within the low-energy bands.\cite{25} Note here that the proposed deformation just models the observed symmetry breaking to provide a proof of principle for the chiral pumping effect. We expect our model to be qualitatively correct regardless of the actual deformations of the lattice close to dislocations.

The effective exchange field seen by itinerant electrons, $\bm{n}\left(t,\mathbf{r}\right)$, is tilted away from the local magnetization following the torsion of the lattice, as illustrated in Fig. 4b. This is implemented by the spin-lattice coupling in Eq.~\eqref{eq:s-l}, which can be understood on geometry grounds as the non-abelian connection that transports the spin quantization axis along the distorted graphene lattice. The details are provided in the Appendix. To the lowest order in $\Delta_{\textrm{so}}$, we obtain\begin{align}
\label{eq:tilting}
\bm{n}\left(t,\mathbf{r}\right)\approx\bm{m}\left(t,\bm{r}\right)-\frac{2a\Delta_{\textrm{so}}}{\hbar v_F}\,\mathbf{b}\left(\mathbf{r}\right)\times\bm{m}\left(t,\mathbf{r}\right),
\end{align}
where $\mathbf{b}\left(\mathbf{r}\right)$ is defined as
\begin{align}
\label{eq:b}
\mathbf{b}\left(\mathbf{r}\right)=a^{-1}\int_{\mathbf{r}_0}^{\mathbf{r}}d\mathbf{r}'\left(\partial_xu_y-\partial_y u_x\right).
\end{align}
Here $\mathbf{r}_0$ is the origin of the dislocation and we have introduced a microscopic length scale $a$ in order to make $\mathbf{b}\left(\mathbf{r}\right)$ dimensionless.

\subsection{Electromotive forces}

The magnetization dynamics induces electromotive forces in SLG along the direction of propagation of spin waves. At low frequencies, majority electrons along the quantization axis defined by $\bm{n}\left(t,\mathbf{r}\right)$ experience an effective electric field of the form\cite{22,23,24}\begin{align}
 E_i=\frac{\hbar}{2e}\dot{\bm{n}}\cdot\left(\bm{n}\times\partial_i\bm{n}+\beta\,\partial_i\bm{n}\right).
 \label{eq:E1}
\end{align}
The first term is purely geometrical, strictly valid in the limit $\Delta_{\textrm{ex}}\gg\hbar \left|\dot{ \bm{n}}\right|$, $\hbar v_F\left|\partial_i \bm{n}\right|$, when the electron spins follow adiabatically the direction of effective exchange field. The $\beta$ correction is related to a slight misalignment due to spin relaxation caused by the spin-orbit interaction.

The electromotive force in the adiabatic limit, which is rooted in the associated geometrical Berry phase, respects the structural symmetries of the device. Indeed, the correction due to the magnetization tilting expressed in Eq.~\eqref{eq:tilting} averages to $0$ when integrated upon the period of the spin wave excitation. The introduction of screw dislocations in YIG breaks the structural symmetries of the transport signals in SLG owing only to non-adiabatic corrections to the electromotive force. These appear as higher order expansions in $\frac{\hbar}{\Delta_{\textrm{ex}}}\dot{\bm{n}}$ and spatial gradients $\frac{\hbar v_F}{\Delta_{\textrm{ex}}}\partial_i\bm{n}$ that capture spin-orbital effects. The new terms compatible with C$_{6v}$ symmetry read
\begin{align}
\label{eq:nonadiabatic}
E_i=\frac{\hbar^3 v_F}{2e\Delta_{\textrm{ex}}^2}\left(\beta_1\,\partial_i\bm{n}+\beta_2\,\bm{n}\times\partial_i\bm{n}\right)\cdot\hat{\mathbf{z}}\times\boldsymbol{\nabla}\left(\dot{\bm{n}}\right)^2,
\end{align}
where $\beta_{1,2}$ are dimensionless phenomenological parameters. The tilting of the exchange field translates the effect of the symmetry breaking in YIG into the electronic response of graphene, driving a voltage along the direction of propagation of the spin waves of the form\cite{footnote}\begin{align}
\label{eq:V}
V=\frac{\beta_1\hbar^2\Delta_{\textrm{so}}}{\Delta_{\textrm{ex}}^2}\int dx \left(\partial_x u_y-\partial_y u_x\right)\hat{\mathbf{z}}\cdot\bm{m}\,\,\partial_x\left(\dot{\bm{m}}\right)^2.
\end{align}
As seen from this equation, the proposed theoretical model reproduces the experimentally observed inversion of the sign of the induced voltage accompanying the reversal of the magnetization ($\hat{\mathbf{z}}\cdot\bm{m}\rightarrow-\hat{\mathbf{z}}\cdot\bm{m}$ when $H_0\rightarrow-H_0$) or the direction of the spin-wave propagation ($\partial_x\left(\dot{\bm{m}}\right)^2\rightarrow-\partial_x\left(\dot{\bm{m}}\right)^2$ when $\bm{k}_{\textrm{sw}}\rightarrow-\bm{k}_{\textrm{sw}}$).

The contribution in Eq.~\eqref{eq:V} should saturate when the theory crosses over to the strongly non-adiabatic regime, at frequencies of the order of $\Delta_{\textrm{ex}}/\hbar$. We expect for the electromotive force to be suppressed at larger frequencies due to a dynamic averaging effect, similarly to the motional narrowing in the spin diffusion problem: the magnetization precession is so fast that its time-dependent component effectively averages out from the view of electron spins. 

\subsection{Frequency dependence}

\begin{figure}
\includegraphics[width=1\columnwidth]{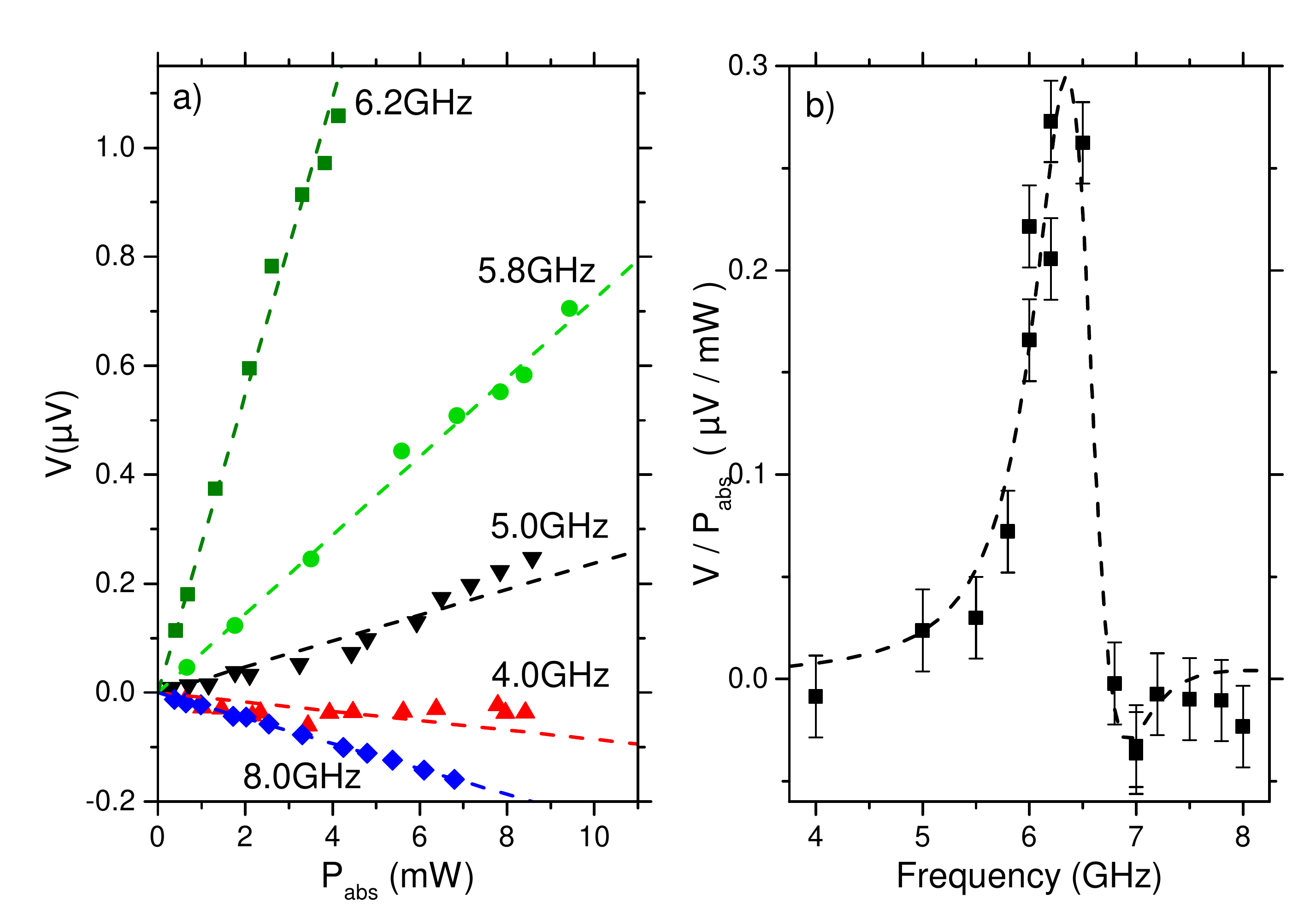}
\caption{Frequency dependence of the effect. a) The detected dc-voltage as a function of the absorbed MW-power for different frequencies as indicated. The dash lines are linear fits of the experimental data used for calculation of the efficiency coefficient $V/P_{abs}$. b) The efficiency coefficient $V/P_{abs}$ as a function of the spin-wave frequency. Note that a sizable effect is observed between 5 and 7 GHz only. The dash line is a guide for the eye.}
\label{Fig5}
\end{figure}

The model predicts that the effect should exist only within a certain range of spin-wave frequencies around $\Delta_{\textrm{ex}}/\hbar$. Although $\Delta_{\textrm{ex}}$ cannot be directly measured in the experiment, the theoretical prediction can be verified by analyzing the frequency dependence of the observed effect. Figure 5(a) shows the dependences of the voltage on $P_{abs}$ proportional to the intensity of the excited spin waves, recorded for different spin-wave frequencies.  As seen from Fig. 5(a), the voltage is linearly proportional to the spin-wave intensity and the proportionality coefficient $V/P_{abs}$ is indeed strongly frequency dependent. Moreover, in agreement with the theoretical results, the frequency dependence of $V/P_{abs}$ (Fig. 5(b)) clearly exhibits a resonance-like behavior with the maximum at about 6 GHz. The relatively small exchange constant $\Delta_{\textrm{ex}}/\hbar\approx30$ eV obtained from the observed resonant frequency nicely agrees with the recently determined exchange field of 0.2 T.\cite{26}

\section{Conclusions}

\label{sec:conclusions}

In conclusion, we experimentally observe a chiral charge pumping in YIG/SLG bilayers caused by the symmetry breaking at the YIG surface. The effect provides a novel, efficient mechanism for detection of magnetization dynamics for spintronic and magnonic applications. The developed theoretical model, taking into account the exchange interaction between localized magnetic moments in YIG and itinerant spins in graphene, predicts that the strength of the effect can be increased by making use of two-dimensional conductors with a strong spin-orbital coupling, such as MoS$_2$.\cite{27}  Additionally, the found symmetry breaking can result in other, not yet observed, effects. In particular, it can enable the reciprocal effect consisting in the excitation of spin-waves of unusual symmetry via an electric current in graphene. These effects provide essentially new opportunities for the direct electrical detection and manipulation of magnetization in insulating magnetic materials and open new horizons for the emerging technologies.

\section*{Acknowledgements}

This work was supported in part by the Deutsche Forschungsgemeinschaft, the Swedish Research Council, the Swedish Foundation for Strategic Research, the Chalmers Area of Advance Nano, the Knut and Alice Wallenberg Foundation, the U.S. Department of Energy, Office of Basic Energy Sciences under Award No. DE-SC0012190 (H.O. and Y.T.), and the program Megagrant No. 14.Z50.31.0025 of the Russian Ministry of Education and Science. J. S. thanks the support of STINT, SOEB, and CTS. S.O.D is thankful to L.A. Melnikovsky for fruitful discussions and to R. Sch\"afer for domain mapping in YIG.

\appendix*

\section{Tilting of the exchange field}

The spin rotational symmetry of graphene Hamiltonian is expressly broken by the spin-lattice coupling in Eq.~\eqref{eq:s-l}, but it can be approximately restored in certain limit by means of a local unitary rotation of the spinor wave function. Such a gauge transformation reads as
\begin{align}
U=\text{Exp}\left[-i\,\zeta\,\mathbf{b}\left(\mathbf{r}\right)\cdot\mathbf{s}\right]\approx\mathcal{I}-i\,\zeta \,\mathbf{b}\left(\mathbf{r}\right)\cdot\mathbf{s},
\label{eq:U}
\end{align}
where $\mathbf{b}\left(\mathbf{r}\right)$ is an axial vector, defined in Eq.~\eqref{eq:b}, and $\zeta$ is a small, dimensionless parameter that we identify with\begin{align}
\zeta=\frac{a\Delta_{\textrm{so}}}{\hbar v_F}\ll1.
\end{align}
Notice that there is a gauge ambiguity in the definition of $U$. In Eq.~\eqref{eq:b}, the gauge is fixed by choosing the origin of the dislocation as the lower limit of integration, but this is arbitrary. As a result, physical effects induced by the tilting of the effective exchange field appear as derivatives of $\mathbf{b}\left(\mathbf{r}\right)$, making this approach fully consistent.

Next, we see that the unitary transformation in Eq.~\eqref{eq:U} gauges away the last term in the Hamiltonian of Eq.~\eqref{eq:H}, generating new terms that are second order in $\zeta$ and we can safely neglect. First, we have that\begin{align}
U^{\dagger}\mathcal{H}_{\textrm{s-l}}U\approx\mathcal{H}_{\textrm{s-l}}-i\zeta\left[\mathcal{H}_{\textrm{s-l}},\mathbf{b}\cdot\mathbf{s}\right]+\mathcal{O}\left(\zeta^2\right),
\end{align}
but note that the strength of $\mathcal{H}_{s-l}$ is already first order in $\zeta$ and therefore the second term in this last equation is actually a second order term; then, we can write\begin{align}
U^{\dagger}\mathcal{H}_{\textrm{s-l}}U\approx\mathcal{H}_{\textrm{s-l}}+\mathcal{O}\left(\zeta^2\right).
\end{align}
On the other hand, if we apply the unitary transformation to the kinetic term in Eq.~\eqref{eq:H} we obtain\begin{align}
& -i\,\hbar\, v_F\, U^{\dagger}\boldsymbol{\Sigma}\cdot\boldsymbol{\partial}U
\approx-\frac{\hbar v_F}{a}\zeta\left(\partial_xu_y-\partial_yu_x\right)\left(\Sigma_x s_x+\Sigma_ys_y\right)
\nonumber\\
& +\mathcal{O}\left(\zeta^2\right)
=-\mathcal{H}_{\textrm{s-l}}+\mathcal{O}\left(\zeta^2\right).
\end{align}
Then, the subsequent gauge field cancels out the spin-lattice coupling. To the leading in $\zeta\ll 1$ (therefore, in the strength of the interaction), the spin-lattice coupling in Eq.~\eqref{eq:s-l} can be understood as the non-abelian connection that transports the electron spin through the distorted crystal, following the torsion created by the screw dislocations in YIG.

We have to apply the transformation to the exchange coupling. Formally, we have\begin{align}
\Delta_{\textrm{ex}}\,U^{\dagger}\,\bm{m}\cdot\mathbf{s}\,U=\Delta_{\textrm{ex}}\,\bm{n}\cdot\mathbf{s},
\end{align}
where $\mathbf{n}=\mathcal{R}\,\mathbf{m}$, and $\mathcal{R}$ is a SO(3) rotation whose matrix elements read\begin{align}
\mathcal{R}_{ij}=\frac{1}{2}\mbox{Tr}\left[s_iU^{\dagger}s_jU\right].
\end{align}
More precisely, $\mathcal{R}$ can be written as\begin{align}
\mathcal{R}=\text{Exp}\left[i\,\zeta\,\mathbf{b}\cdot\boldsymbol{\ell}\right],
\end{align}
where the SO(3) generators $\boldsymbol{\ell}=\left(\ell_x,\ell_y,\ell_z\right)$ in this basis read
\begin{gather*}
\ell_x=\left(\begin{array}{ccc}
0 & 0 & 0\\
0 & 0 & -2i\\
0 & 2i & 0
\end{array}\right),\nonumber\\
\ell_y=\left(\begin{array}{ccc}
0 & 0 & 2i\\
0 & 0 & 0\\
-2i & 0 & 0
\end{array}\right),\nonumber\\
\ell_z=\left(\begin{array}{ccc}
0 & -2i & 0\\
2i & 0 & 0\\
0 & 0 & 0
\end{array}\right).
\end{gather*}
To the leading order in $\zeta$ we have\cite{note}\begin{align}
\bm{n}\approx\bm{m}-2\text{ }\zeta\text{ }\mathbf{b}\times\bm{m}+\mathcal{O}\left(\zeta^2\right),
\end{align}
which corresponds to Eq.~\eqref{eq:tilting}.


\begin{thebibliography}{60}


\bibitem{pumping_theo1} Y. Tserkovnyak, A. Brataas, and G. E. W. Bauer, Phys. Rev. Lett. \textbf{88}, 117601 (2002).

\bibitem{pumping_theo2} A. Brataas, Y. Tserkovnyak, G. E. W. Bauer, and B. I. Halperin, Phys. Rev. B \textbf{66}, 060404(R) (2002).

\bibitem{mismatch} G. Schmidt, D. Ferrand, L. W. Mollenkamp, A. T. Filip, B. J. van Wees, Phys. Rev. B \textbf{62}, R4790(R) (2000).

\bibitem{pumping_exp1} Bret Heinrich, Yaroslav Tserkovnyak, Georg Woltersdorf, Arne Brataas, Radovan Urban, and Gerrit E. W. Bauer, Phys. Rev. Lett. \textbf{90}, 187601 (2003).

\bibitem{pumping_exp2} K. Lenz, T. Toli\'nski, J. Lindner, E. Kosubek, and K. Baberschke, Phys. Rev. B \textbf{69}, 144422 (2004).

\bibitem{pumping_exp3} M. V. Costache, M. Sladkov, S. M. Watts, C. H. van der Wal, and B. J. van Wees, Phys. Rev. Lett. \textbf{97}, 216603 (2006).

\bibitem{spin-charge1} E. Saitoh, M. Ueda, H. Miyajima, and G. Tatara, Appl. Phys. Lett. \textbf{88}, 182509 (2006).

\bibitem{spin-charge2} O. Mosendz, J. E. Pearson, F. Y. Fradin, G. E. W. Bauer, S. D. Bader, and A. Hoffmann, Phys. Rev. Lett. \textbf{104} 046601 (2010).

\bibitem{bilayer} A. Azevedo, L. H. Vilela-Le\~ao, R. L. Rodr\'iguez-Su\'arez, A. F. Lacerda Santos, and S. M. Rezende, Phys. Rev. B \textbf{83}, 144402 (2011).

\bibitem{10} K. S. Novoselov, A. K. Geim, S. V. Morozov, D. Jiang, Y. Zhang, S. V. Dubonos, I. V. Grigorieva, and A. A. Firsov, Science \textbf{306}, 666 (2004).
\bibitem{12} A. H. Castro Neto, F. Guinea, N. M. R. Peres, K. S. Novoselov, and A. K. Geim, Rev. Mod. Phys. \textbf{81}, 109 (2009).
\bibitem{6} S. K. Keun \textit{et al}., Nature \textbf{457}, 706 (2009).
\bibitem{7} X. Li \textit{et al}., Science \textbf{324}, 1312 (2009).
\bibitem{8} S. Bae \textit{et al}., Nature Nanotechnology \textbf{5}, 574 (2010).
\bibitem{13} V. Cherepanov, I. Kolokolov, and V. L'vov, Phys. Rep. \textbf{229}, 81 (1993).
\bibitem{14} A. A. Serga, A. V. Chumak, and B. Hillebrands, J. Phys. D \textbf{43}, 264002 (2010).
\bibitem{15} Y. Kajiwara \textit{et al}., Nature \textbf{464}, 262 (2010).
\bibitem{16} A. Hamadeh \textit{et al}, Phys. Rev. Lett. \textbf{113}, 197203 (2014).
\bibitem{17} Zhenyao Tang, Eiji Shikoh, Hiroki Ago, Kenji Kawahara, Yuichiro Ando, Teruya Shinjo, and Masashi Shiraishi, Phys. Rev. B \textbf{87}, 140401(R) (2013).
\bibitem{9} J. B. S. Mendes, O. Alves Santos, L. M. Meireles, R. G. Lacerda, L. H. Vilela-Le\~ao, F. L. A. Machado, R. L. Rodr\'iguez-Su\'arez, A. Azevedo, and S. M. Rezende, Phys. Rev.  Lett. \textbf{115}, 226601 (2015).
\bibitem{18} S. Singh, A. Ahmadi, C. T. Cherian, E. R. Mucciolo, E. del Barco, and B. \"Ozyilmaz, Appl. Phys. Lett. \textbf{106}, 032411 (2015).
\bibitem{28} J. Sun, \textit{et al}., IEEE Trans. Nanotechnol. \textbf{11}, 255 (2012).
\bibitem{29} L. Gao, \textit{et al}., Nat Commun. \textbf{3}, 699 (2012).
\bibitem{30} C. J. Lockhart de la Rosa, \textit{et al}., Appl. Phys. Lett. \textbf{102}, 022101 (2013).
\bibitem{19} S. A. Nikitov, \textit{Relaxation Phenomena of Magnetic Excitations in Ferromagnetic Media}, in \textit{Advances in Chemical Physics: Relaxation Phenomena in Condensed Matter}, Vol. 87, ed. by W. Coffey (John Wiley, New Jersey, 1994).
\bibitem{20} P. Nowik-Boltyk, O. Dzyapko, V. E. Demidov, N. G. Berloff, and S. O. Demokritov, Sci. Rep. \textbf{2}, 482 (2012).
\bibitem{21} Ioan Mihai Miron, Gilles Gaudin, St\'ephane Auffret, Bernard Rodmacq, Alain Schuhl, Stefania Pizzini, Jan Vogel, and Pietro Gambardella, Nat. Mater. \textbf{9}, 230 (2010).
\bibitem{25} H. Ochoa, A. H. Castro Neto, V. I. Fal'ko, and F. Guinea, Phys. Rev. B \textbf{86}, 245411 (2012).
\bibitem{22} G. E. Volovik, J. Phys. C: Solid State Phys. \textbf{20}, L83 (1987).
\bibitem{23} R. A. Duine, Phys. Rev. B \textbf{77}, 014409 (2008).
\bibitem{24} Y. Tserkovnyak and M. Mecklenburg, Phys. Rev. B \textbf{77}, 134407 (2008).
\bibitem{footnote} By plugging Eq.~\eqref{eq:tilting} into Eq.~\eqref{eq:nonadiabatic}, the $\beta_1$ term generates the longitudinal voltage in Eq.~\eqref{eq:V}, whereas the $\beta_2$ term leads to a transverse voltage. The latter is even in the applied magnetic field, in agreement with symmetry arguments. This prediction is in fact supported by preliminary experiments. A careful study of the induced transverse voltage will be a topic our further investigations.
\bibitem{26} C. Leutenantsmeyer, A. A. Kaverzin, M. Wojtaszek, and B. J. van Wees, arXiv:1601.00995 [cond-mat.mes-hall].
\bibitem{27} K. F. Mak, C. Lee, J. Hone, J. Shan, and T. F. Heinz, Phys. Rev. Lett. \textbf{105}, 36805 (2010).

\bibitem{note} The action of the SO(3) generators over a vector $\mathbf{v}$ satisfy the following property: $\left(\mathbf{u}\cdot\boldsymbol{\ell}\right)\,\mathbf{v}=2\,i\,\mathbf{u}\times\mathbf{v}$.

\end{thebibliography}
\end{document}